\documentclass[useAMS,usenatbib]{mn2e}

\usepackage{aas_macros}
\usepackage{times}
\usepackage{graphics,epsfig}
\usepackage{graphicx}
\usepackage{amsmath}
\usepackage{amssymb}
\usepackage[breaklinks,colorlinks=true,citecolor=black,linkcolor=black,urlcolor=blue,bookmarks=true]{hyperref}

\usepackage{ulem}

\newcommand{\kms}{km\,s$^{-1}$}
\newcommand{\Lsun}{$L_{\sun}$}
\newcommand{\Msun}{$M_{\sun}$}
\newcommand{\MLsun}{\Msun{}\,\Lsun$^{-1}$}
\newcommand{\SFR}{\Msun{}\,yr$^{-1}$}
\newcommand{\SSFR}{\Msun{}\,yr$^{-1}$\,\Lsun$^{-1}$}
\newcommand{\HI}{\textrm{H}\,\textsc{i}}

\title[A привет coupling pair of dwarfs in Lynx]{A coupling pair of dwarfs in Lynx}
\author[Makarov et al.]{
Dmitry Makarov$^{1,2}$\thanks{E-mail: dim@sao.ru},
Igor D.\ Karachentsev$^{1,2}$,
Jayaram N.\ Chengalur$^3$,
\newauthor
Roman Uklein$^1$
and
Alexander Marchuk$^4$
\\
$^1$ Special Astrophysical Observatory, Nizhniy Arkhyz, Karachai-Cherkessia 369167, Russia\\
$^2$ Leibniz-Institut f\"{u}r Astrophysik (AIP), An der Sternwarte 16, D-14482 Potsdam, Germany\\
$^3$ National Centre for Radio Astrophysics, TIFR, Ganeshkhind, Pune 411 007, India\\
$^4$ St Petersburg State University, Universitetskii pr.\ 28, 198504 St Petersburg, Stary Peterhof, Russia
}

\begin{document}
\date{}

\pagerange{\pageref{firstpage}--\pageref{lastpage}} \pubyear{2013}

\maketitle

\label{firstpage}

\begin{abstract}
We report on discovery of unique binary system of dIr galaxies 
which looks like a low surface brightness tidal stream in the halo of another dwarf galaxy. 
Both the galaxies are detected in the \HI{} line with the Giant Metrewave Radio Telescope (GMRT). 
The pair components have absolute blue magnitudes $-16.0$ and $-13.2$, 
heliocentric radial velocities 1799 km/s and 1842 \kms{}, 
and hydrogen masses-to-luminosity ratios 0.4 and 1.6 in solar units, respectively. 
The binary system is characterized by a high orbital mass-to-luminosity ratio of 49 \MLsun{} 
and a short crossing time of 0.22 Gyr.
\end{abstract}

\begin{keywords}
galaxies: dwarf -- radio lines: galaxies
\end{keywords}

\section{Introduction}

In the current paradigm of structure formation in the Universe \citep{WR1978}, 
galaxies are built from the merging of small units.
Recent studies have found substantial evidence for an ongoing 
major (i.e.\ the progenitor galaxies have similar masses) and 
minor (i.e.\ the progenitor galaxies have significantly different masses, small satellite merging with a large host) 
mergers in nearby massive galaxies \citep[see][]{M+2009,MD+2010}. 
In the hierarchical scenario even small galaxies are expected to be formed by merging of even smaller objects. 
However, still now there is little direct evidence of this and 
the number of studies of interacting dwarf systems are very limited.

Recently, \citet{MD+2012} and \citet{R+2012} studied the tidally distorted dwarf galaxy d1228+4348
situated near the star bursting dwarf Im-type galaxy NGC\,4449.
The very low surface brightness (LSB) object d1228+4358 was originally discovered by \citet{KKH2007}.
The nearby dwarf NGC\,4449 with $M_V \sim -18.6$ at distance of 4.21 Mpc is similar to the LMC, but has a much higher star formation rate. 
Earlier, \HI{} morphology of NGC\,4449 reveals existence of large, highly structured, extended clouds and very long streamers, 
which are a consequence of interaction, but without clearly identified disturbing galaxy \citep{HWE1998}.
The fresh data show that the mass ratio of the two galaxies is extreme high ($\sim 1:50$), 
indicating that even very minor mergers could have a profound effect on the interstellar medium and the star formation.
Here we report on an optical and \HI{} properties of new, very unusual object found near the dwarf galaxy SDSS\,J091108.40+423922.1.

\section{Optical properties of the dwarfs}

\begin{figure}
\psfig{file=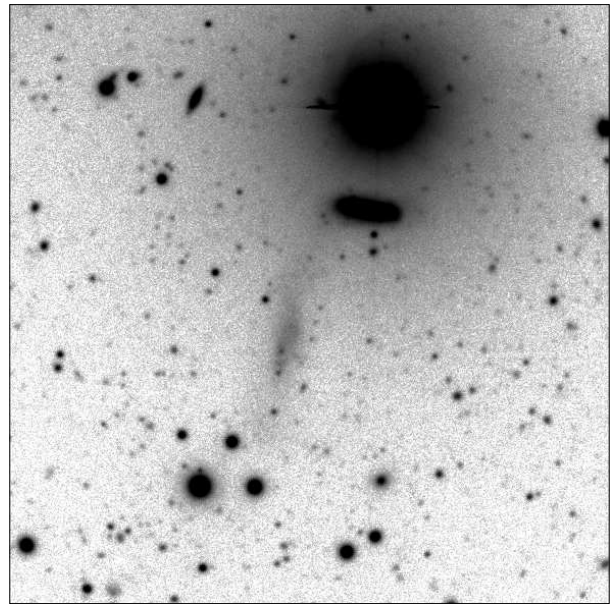,width=0.45\textwidth,angle=0}
\caption{
$V$-band direct image of the LSB\,J0911+4238 system in logarithmic scale in grey colormap.
It was received with 6-meter SAO RAS telescope.
}
\label{fig:bta}
\end{figure}

A low surface brightness object (henceforth LSB\,J0911+4238) was found by Dmitry Makarov near the dwarf galaxy SDSS\,J091108.40+423922.1 
during the process of morphological classification of $\sim 11,000$ galaxies in the Local Universe with radial velocities less than 3500 \kms{}. 
The pair J0911+42 was included in a list of binary and multiple dwarf galaxies by \citet{MU2012}.
The bright SDSS\,J091108.40+423922.1 component of the pair is J0911+42A,
and LSB\,J0911+4238 has a designation J0911+42B.
The LSB object is barely seen on the DSS and SDSS images.
In March 2012 it was imaged in a $V$-band with the 6-meter telescope 
of the Special Astrophysical Observatory of the Russian Academy of Sciences (SAO RAS) (Fig.~\ref{fig:bta}). 
The bright SDSS galaxy has a heliocentric velocity of $1818\pm10$ \kms{} obtained by emission lines from SDSS spectral data archive.
It has blue colour and substantial FUV emission (with a magnitude of 17.43) 
consistent with the significant ongoing star formation, and hence reasonable \HI{} gas content.
The neighbouring object LSB\,J0911+4238 also has blue colour, suggesting that it is gas rich too. 
We carried out surface photometry of both components, using SDSS DR9 data.
Asymptotic total magnitudes were estimated from aperture photometry in ellipses, 
$g=15.69$, $r=15.40$, $i=15.29$ for bright component, and 
$g=18.54$, $r=18.21$, $i=18.19$ for LSB galaxy.
We estimated $B=16.02$ for A, and $B=18.88$ for B-component, 
using transformation of SDSS magnitudes to standard Johnson system by \citet{SDSSphotometry}.
For a kinematic distance of $V/73 = 25$ Mpc, the galaxies have an absolute B-magnitude of $-16.0$ and $-13.2$ mag, respectively.
The linear separation between the SDSS galaxy and the newly discovered LSB object is $1.35\arcmin$, which corresponds to $\sim 10$~kpc.  
The basic parameters of the galaxies are presented in Table~\ref{tab:main}.

\begin{table*}
\caption{Basic properties of the pair of dwarfs}
\label{tab:main}
\begin{tabular}{lccc}
\hline\hline
                                     & SDSS\,J091108.40+423922.1 &  LSB\,J0911+4238   &   Pair         \\
\citet{MU2012} designation                  & J0911+42A          &  J0911+42B         &   J0911+42     \\
J2000                                & 09\,11\,08.4 +42\,39\,22  & 09\,11\,10.6 +42\,38\,01 &          \\
\hline
Angular dimension/separation, arcmin        &$ 0.64 \times 0.23 $&$  1.0 \times 0.3  $&$   1.35       $\\
Apparent B-magnitude, mag                   &$ 16.02            $&$  18.88           $&$              $\\
Apparent FUV magnitude, mag                 &$ 17.43 \pm0.05    $&$   >23.0          $&$              $\\
\HI{} flux, Jy\,\kms{}                      &$ 1.15 \pm 0.1     $&$  0.32 \pm 0.05   $&$ 1.53 \pm 0.1 $\\
\HI{} linewidth, \kms{}                     &$   57.3\pm4       $&$    34.1\pm6      $&$   68.4\pm4   $\\
Heliocentric optical velocity, \kms{}       &$ 1818\pm10\;^{*)} $&$                  $&$              $\\
Heliocentric \HI{} velocity, \kms{}         &$ 1799.3\pm3       $&$  1841.6\pm5      $&$ 1811.1\pm1   $\\
Linear diameters/separation, kpc            &$  4.6 \times 1.7  $&$   7.2 \times 2.2 $&$   9.7        $\\       
Absolute B-magnitude, mag                   &$   -16.01         $&$    -13.15        $&$              $\\
Blue luminosity, $\times10^8$ \Lsun         &$  3.94            $&$   0.28           $&$  4.23        $\\
\HI{} mass, $\times10^8$ \Msun              &$  1.65            $&$   0.46           $&$  2.20        $\\
Dynamical/orbital mass, $\times10^8$ \Msun  &$  4.39            $&$   2.43           $&$  206         $\\
$\log(\textrm{SSFR})$, \SSFR                &$  -9.95           $&$   <-11.0         $&$              $\\
$\log(\textrm{SFR})$, \SFR                  &$  -1.35           $&$   < -3.58        $&$              $\\
\hline
\multicolumn{4}{p{0.7\textwidth}}{
$^{*)}$ SDSS redshift estimation before DR8 gives $V_h = 1496 \pm13$ \kms{} from cross-correlation templates.  
This value significantly underestimates the real redshift of the galaxy.
}\\
\hline\hline
\end{tabular}
\end{table*}

One can put forward various possibilities to explain properties of the pair from the available data.

\begin{enumerate}
\item
The LSB object and the SDSS dwarf are spatially disconnected objects seen in projection by chance. 
However, this is unlikely, because of close proximity of these two faint dwarfs.
Further, from the optical images one gets the impression that these two objects are in contact each other.

\item 
We see a case of merging the two dwarf galaxies. 
However, it looks rather unusual that the dwarf galaxy of $-16.0$ mag produces such strong distortion of shape of the LSB dwarf. 
No other similar examples are known so far. 

\item
The LSB object is an usual tidal tail of the SDSS galaxy produced by its interaction with neighbouring galaxy.
However, the SDSS dwarf is a rather isolated one. 
It resides in far outskirts of NGC\,2798 group, which contains 6 members with the mean heliocentric velocity of 1704 \kms{}, 
velocity dispersion of 73 \kms{} and harmonic radius of 75 kpc \citep{Groups}. 
The SDSS galaxy is at a projected distance of 560 kpc from the brightest group members NGC\,2798/99 = Arp283 = VV50 = KPG195.
   
\item 
It may be a case of interaction of the dwarf SDSS galaxy with a dark invisible halo.
Thus, the LSB object is a tidal tail of the SDSS.
A dozen of very isolated galaxies, like UGC\,4722, with tails or other shape distortions have been discussed by \citet{KIAUS244}.

\end{enumerate}

Having not sufficient observational data to clearly distinguish between the various scenarios proposed above, 
we undertook \HI{} imaging of this unique and puzzling system to provide a crucial input in its understanding.

\section{\HI{} observations}
\label{sec:HIobs}
 
\HI{} 21 cm observations were conducted using the Giant Metrewave Radio Telescope (GMRT) on September 14, 2012. 
The total on source time was 1.8 hours. 
The observations had a total bandwidth of 4.16 MHz starting at 1409.77 MHz, or a velocity range from 1392 \kms{} to 2282 \kms{}. 
There were a total of 512 spectral channels, giving a channel width of 1.738 \kms{}. 
The initial flagging and calibration was carried out using the FLAGCAL pipeline \citep{FLAGCAL} after which imaging was done in AIPS. 
The imaging was performed after boxcar smoothing by 4 channels, i.e.{} to a velocity resolution of 6.95 \kms{}. 
A continuum image was made using the line free channels, and subtracted from the visibilities using the task UVSUB. 
After this image cubes were made using the task IMAGR, and moment maps using the task MOMNT.

\begin{figure}
\psfig{file=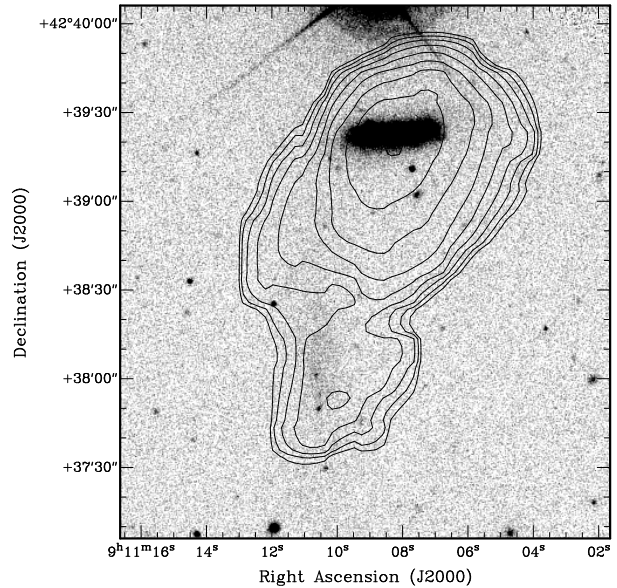,width=0.45\textwidth,angle=0}
\caption{
GMRT integrated \HI{} image (contours) of the LSB\,J0911+4238 system, overlayed on the SDSS g band image. 
The angular resolution of the image is 52\arcsec.
The contour levels start at $2 \times 10^{19}$ atoms cm$^{-2}$ and increase in steps of $\sqrt{2}$.
}
\label{fig:hi52m0}
\end{figure}

\begin{figure}
\psfig{file=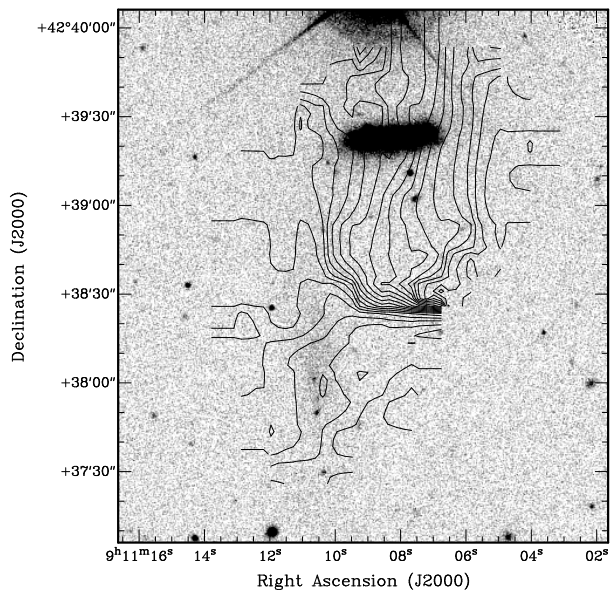,width=0.45\textwidth,angle=0}
\caption{
GMRT \HI{} velocity field (contours) of the LSB\,J0911+4238 system, overlayed on the SDSS g band image.
The angular resolution of the image is 52\arcsec.
The contour levels start at 1744~\kms{} and increase in steps of 3~\kms{}.
}
\label{fig:hi52m1}
\end{figure}

\begin{figure}
\psfig{file=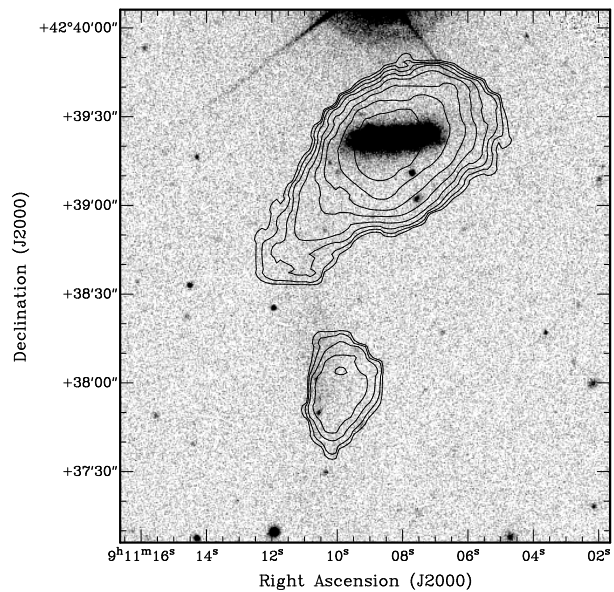,width=0.45\textwidth,angle=0}
\caption{
GMRT integrated \HI{} image (contours) of the LSB\,J0911+4238 system, overlayed on the SDSS g band image. 
The angular resolution of the image is 35\arcsec.
The contour levels start at $4 \times 10^{19}$ atoms cm$^{-2}$ and increase in steps of $\sqrt{2}$.
}
\label{fig:hi35m0}
\end{figure}

In Fig.~\ref{fig:hi52m0} is shown an overlay of the integrated \HI{} emission (contours) on the SDSS g-band image. 
The resolution of the \HI{} emission is 52\arcsec{}. 
Emission is clearly detected from both SDSS\,J091108.40+423922.1 as well as the companion galaxy LSB\,J0911+4238. 
They are joined by a bridge of \HI{} emission. 
Fig.~\ref{fig:hi52m1} shows the \HI{} velocity field at the same angular resolution. 
One can see a smooth variation of velocity starting from LSB\,J0911+4238 through the bridge and onto SDSS\,J091108.40+423922.1.
Fig.~\ref{fig:hi35m0} presents the integrated \HI{} at a slightly higher angular resolution, 35\arcsec{}. 
The bridge emission is partly resolved out, but one can clearly see \HI{} associated with LSB\,J0911+4238, 
as well as a tidal extension to the \HI{} associated with SDSS\,J091108.40+423922.1.

\begin{figure}
\psfig{file=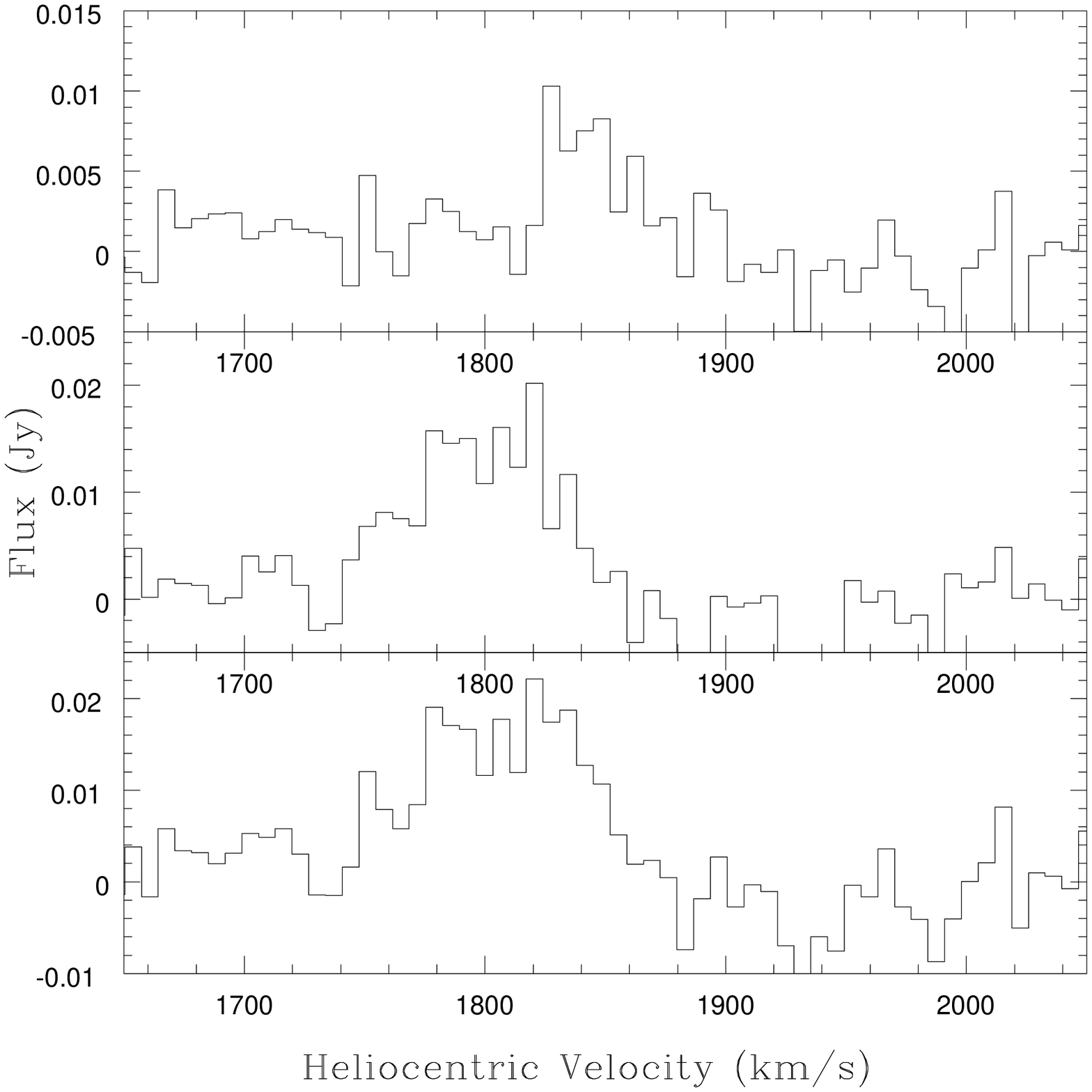,height=3.5truein,angle=0}
\caption{
The integrated \HI{} spectra for LSB\,J0911+4238 (top panel),
SDSS\,J091108.40+423922.1 (middle panel), 
and the LSB\,J0911+4238 system as a whole (bottom panel).
}
\label{fig:hispectra}
\end{figure}

The integrated spectra of LSB\,J0911+4238, SDSS\,J091108.40+423922.1, as well as the entire system are shown in Fig.~\ref{fig:hispectra}. 
The central velocities and FWHM of the spectra as obtained from Gaussian fits to the spectra are 
$1841.6\pm5$ \kms{}, $34.1\pm6$ \kms{} for LSB\,J0911+4238, 
$1799.3\pm3$ \kms{}, $57.3\pm4$ \kms{} for SDSS\,J091108.40+423922.1, 
and $1811.1\pm1$ \kms{}, $68.4\pm4$ \kms{} for the system as a whole. 
The \HI{} fluxes for them measured from the spectra are 
$0.32\pm0.05$ Jy\,\kms{}, $1.15\pm0.1$ Jy\,\kms{}, and $1.53\pm0.1$ Jy\,\kms{}, respectively.

\section{Global parameters of the pair of dwarfs}

Table~\ref{tab:main} presents some basic properties of components of this unusual galaxy pair.
To estimate \HI{} galaxy mass we used a standard relation $M(\HI) = 2.36\times10^5 D^2 F(\HI)$,
where \HI{} mass expressed in solar masses, the flux $F(\HI)$ in Jy\,\kms{} and distance $D$ in Mpc. 
As seen, the total \HI{} mass-to-blue luminosity ratio is moderate and 
equal to 0.42 and 1.6 in solar units for bright and faint components, respectively. 
We also estimated a dynamical mass of each component as $M_{\rm dyn} = 2.33\times10^5 (W_{50}/2)^2 A/2$, 
where the $W_{50}$ is a \HI{} linewidth at half of the maxima in \kms{}, $A$ means galaxy linear diameter in kpc. 
The dynamical mass-to-luminosity ratio turns out to be 1.1 (SDSS) and 8.4 (LSB), which is typical for late-type dwarf irregulars.
Note, that mass estimation for LSB\,J0911+4238 is probably highly overestimated because of tidal distortion.

Knowing the radial velocity difference for the components, $\Delta V = 42.3\pm6$ \kms{} and their projected separation, $R_p = 9.7$ kpc, 
we estimated the projected (orbital) mass of the binary system via relation  $M_p = 1.18\times10^6 \Delta V^2 R_p$ \citep{HTB1985}.
It yields us $M_p = 2.06\times10^{10}$ \Msun{}, i.e.\ 30 times higher than the sum of dynamical mass of the pair components.
This quantity as well as the orbital mass-to-total luminosity ratio $49$ \MLsun{} indicates a significant amount of dark matter in the pair. 
Recently, \citet{M+2010} discovered a very faint \HI-rich companion AGES\,J030039+254656 in the vicinity of isolated BCD galaxy NGC\,1156. 
Considering them as a physical pair of dwarfs with $\Delta V = 77\pm4$ \kms{} and $R_p = 80$ kpc
\citet{M+2010} derived its minimal dynamical mass-to-B-luminosity ratio $\sim65$ in solar units. 
There are some indications that isolated pairs of dwarfs have 2--3 times higher orbital mass-to-luminosity ration 
than that of pairs of normal spirals \citep{Pairs}.

Last two lines in Table~\ref{tab:main} present the total and the specific star formation rate for the galaxies
estimated as $\log[SFR] = 2.78 - 0.4 m_{\rm FUV} + 2 \log D$ via far UV magnitudes from GALEX observations \citep{GALEXatlas}. 
As one can see, the star formation rate of SDSS component seems to be typical for dIrs, 
and the upper limit of SSFR for LSB one can be classified as depressed.

Note that a crossing time $t_{\rm cr} = R_p/\Delta V$ for the considered pair is only 0.22 Gyr 
meaning that the binary dwarfs will merged each other very shortly.

\section*{Acknowledgements} 
We are thankful to Alexei Moiseev who provided us with the $V$-band image of LSB\,J0911+4238 obtained with the 6-m telescope of SAO RAS. 
This work was supported by RFBR grants 11--02--00639, RFBR-DFG grant 12--02--91338 and Russian-Indian RFBR grant 13--02--92690.
We acknowledge the support of the Ministry of Education and Science of the Russian Federation, the proposal 2012--1.5--12--000--1011--004.

\bibliographystyle{mn2e}
\bibliography{j0911+42}   

\bsp
\label{lastpage}

\end{document}